\documentclass{apex}

\title{General Rule and Materials Design of Negative Effective $U$ System for High-$T_c$ Superconductivity}

\author{Hiroshi Katayama-Yoshida\thanks{E-mail address: hiroshi@mp.es.osaka-u.ac.jp}, Koichi Kusakabe, Hidetoshi Kizaki, and Akitaka Nakanishi}
\inst{Graduate School of Engineering Science, Osaka University, Toyonaka, Osaka 567-8531, Japan}
\abst{Based on the microscopic mechanisms of (1) charge-excitation-induced negative effective $U$ in $s^1$ or $d^9$ electronic configurations, and (2) exchange-correlation-induced negative effective $U$ in $d^4$ or $d^6$ electronic configurations, we propose a general rule and materials design of negative effective $U$ system in itinerant (ionic and metallic) system for the realization of high-$T_c$ superconductors.  We design a $T_c$-enhancing layer (or clusters) of charge-excitation-induced negative effective $U$ connecting the superconducting layers for the realistic systems.}

\begin{document}
\maketitle

In order to realize a new-class of high-$T_c$ superconductors, 
we need to design and realize an itinerant (ionic and metallic) 
system carrying a large negative effective-correlation energy 
$U$ (NEU) with $U= E(N+1) + E(N-1) - 2E(N) < 0$, 
where $E(N)$ is a total energy of an $N$ electron system.  
In addition to both 
Anderson's mechanism for a negative effective $U$ mechanism~\cite{1} 
induced by Jahn-Teller lattice distortion 
and the exchange-correlation-induced negative effective $U$ (ECI-NEU) 
mechanism proposed by Katayama-Yoshida and Zunger~\cite{2}, here, 
we propose a new universal class of microscopic mechanism 
of NEU caused by charge-excitation-induced NEU (CEI-NEU)
from $s$- to $p$-orbital, or, $d$- to $s$-orbital in an ionic metal.  
The NEU interaction between two electrons generally 
leads to a charge-density wave (CDW) by charge disproportionation, 
or superconductivity (SC) by attractive pairing-interaction, 
or a spin-density wave (SDW) 
by an exchange-correlation interaction.  
In this paper, we propose a general rule on chemical trends for 
NEU system in order to design 
a new-class of high-$T_c$ superconductor, 
which is always competing with a CDW or SDW state. 

The total energy $E(N)$ of $N$ electron system shows normally 
a convexity as a function of $N$ due to the repulsive correlation energy of 
$U$ ($U= E(N+1) + E(N-1) - 2E(N)>0$) between two electrons, 
therefore $U>0$ guarantees a stability 
of the $N$ electron system in the thermal equilibrium (see Fig.1). 
As is depicted schematically in Fig. 1, 
if $(N-1)$ electron system in $s^0$ (or $d^8$, or $d^3$) 
electronic configuration 
or $(N+1)$ electron system in $s^2$ (or $d^{10}$, or $d^5$) 
electronic configuration
becomes more stable than $N$ electron system in $s^1$ (or $d^9$, or $d^4$) 
electronic configuration by lowering the total energy 
in the $(N+1)$ electron system through mixing of $s^2$ and 
charge excited states with $p$ electrons 
(or $d^{10}$ and charge excited states with $s$ electrons), 
or through the exchange-energy gain of $d^5$ electronic configuration, 
the $N$ electron system in $s^1$ (or $d^9$, or $d^4$) electronic configuration 
becomes NEU system due to the concavity (see Fig.~1).  
In {\it ab initio} calculation, NEU system caused by 
ECI-NEU mechanism is predicted 
in Si:Cr system~\cite{2} and in (Ga,Mn)As system~\cite{5} quantitatively. 
We can realize NEU system 
for $s^1$ (or $d^9$) electronic configuration based on 
the charge excitation or for $d^4$ electronic configuration 
based on the exchange-correlation energy gain~\cite{2} 
as will be shown later, 
therefore, we can expect a charge disproportionation 
and insulating CDW in the reaction of $2s^1 \rightarrow s^0 + s^2$, 
(or $2d^9 \rightarrow d^8 + d^{10}$, or $2d^4 \rightarrow d^3 + d^5$).  
However, if we can realize an itinerant 
NEU system upon $p$- or $n$-type doping or 
under the applied ultra-high pressures by destabilizing CDW or SDW, 
we can stabilize a superconducting state caused 
by electron pairing through attractive-carrier dynamics 
in CEI-NEU system.

Two important pure electronic mechanisms of NEU are 
(i) the present microscopic (electronic) mechanism of 
NEU system in the $s^1$ or $d^9$ electronic configurations 
by stabilizing the electron-rich $s^2$ or $d^{10}$ electronic configurations 
through the charge excitation from 
the $s$- to $p$-orbital or from the $d$- to $s$-orbital 
and (ii) ECI-NEU~\cite{2} 
in $d^4$ (or $d$ holes in $d^6$) electronic configuration due to 
the exchange-correlation stabilized $d^5$ electronic configuration 
(Hund's rule). 

Indication of NEU system due to these pure electronic mechanisms 
appears as the missing oxidation states in experiments. 
ECI-NEU causes 
the missing oxidation state in the $d^4$ (or $d^6$) electronic configuration 
through the charge disproportionations 
in the reaction of $2d^4 \rightarrow d^3 + d^5$, 
(or, $2d^6 \rightarrow d^5 + d^7$). 
Katayama-Yoshida and Zunger~\cite{2} proposed 
that localized centers sustaining local magnetic moment 
can show NEU behavior, 
when the exchange (or, in general, many-electron correlation) 
interactions outweigh the strongly reduced Coulomb repulsions 
in the covalent materials such as oxides or covalent semiconductors. 
This NEU mechanism can explain 
missing oxidation states in chemistry~\cite{3}; 
e.g., while both Mn$^{2+}$ ($d^5$) 
and Mn$^{4+}$($d^3$) are observed in MgO:Mn and CaO:Mn, 
the Mn$^{3+}$($d^4$) center is missing~\cite{4} ; 
also in (Ga,Mn)As system Mn$^{3+}$($d^4$) center is missing~\cite{5}.  
If we can realize itinerant system 
with ECI-NEU system 
with destabilizing the ferromagnetic SDW caused by Zener's double exchange 
or Zener's $p$-$d$ exchange mechanisms~\cite{6,7}, 
we have a possibility to stabilize the superconducting state 
in $d^4$ (or $d^6$) electronic configuration 
by ECI-NEU.  
However, SDW is generally more stable than a superconducting state 
in Mn$^{3+}$($d^4$) with large exchange-correlation energy gain 
in (Ga,Mn)As or perovskite manganites.

We have another microscopic mechanism 
of NEU system in the $s^1$ or $d^9$ electronic configurations. 
Here we may refer to the experimentally observed chemical trends of first 
($A^0 \rightarrow A^+ + e^-$), second ($A^+ \rightarrow A^{2+} + e^-$), 
and third ($A^{2+} \rightarrow A^{3+} + e^-$) ionization energies 
of free atoms in the second, third, fourth, and fifth period 
in the periodic table.  
We find that the ionization energy of $s^1$ electronic configuration 
for $2s$-, $3s$-, $4s$-, and $5s$-orbitals shows the minimum 
suggesting an instability of $s^1$ electronic configuration, 
and that $s^2$ electronic configurations (closed-shell in $s$-orbital) 
show always the peak 
suggesting a stability of closed-shell in $s^2$ 
electronic configuration~\cite{Martienssen}.

The experimental data on the missing oxidation states of ions 
in $s^1$ electronic configuration may be 
another secret key point. 
It is well known in experiment that the existence of 
the missing oxidation states of ions such as 
Tl$^{2+}$ ($6s^1$), Pb$^{3+}$($6s^1$), Hg$^+$($6s^1$), and Bi$^{4+}$($6s^1$) 
in the periodic table~\cite{19,20}, and these atoms indicate 
NEU nature, where following charge 
disproportionation occurs in the insulating compounds; 
2Tl$^{2+}$($6s^1$) $\rightarrow$ Tl$^+$($6s^2$) $+$ Tl$^{3+}$($6s^0$), 
2Pb$^{3+}$($6s^1$) $\rightarrow$ Pb$^{2+}$($6s^2$) $+$ Pb$^{4+}$($6s^0$), 
2Hg$^+$($6s^1$) $\rightarrow$ Hg$^0$($6s^2$) $+$ Hg$^{2+}$($6s^0$), and 
2Bi$^{4+}$($6s^1$) $\rightarrow$ Bi$^{3+}$($6s^2$) $+$ Bi$^{5+}$($6s^0$).  
Therefore, all of these insulating compounds such as BaBiO$_3$, BaPbO$_3$, 
indicate the insulating CDW with static charge disproportionation 
caused by CEI-NEU.  
In order to realize a new-class of high-$T_c$ superconductivity 
based on CEI-NEU 
in an itinerant system, 
we should destabilize CDW by $p$- or $n$-type doping or 
by applying the ultra-high pressures. 

We show our theoretical prediction of missing oxidation states 
comparing with the experimental observations~\cite{19, 20}, 
where all of the electronic configurations in the missing oxidation states 
are the $s^1$ electronic configuration acting as a NEU center 
(see Table I).
The microscopic mechanism of the NEU system 
in the $N$ electron system of $s^1$ electronic configuration 
in the polarizable condensed matter, 
such as solid, liquid or glass materials, is caused by 
the closed-shell stability in the ($N+1$) electron system of $s^2$ electronic 
configuration due to the short-range charge excitation from 
the $s$- to $p$-orbital in condensed matter.  
Since the energy level of the ground state in the $ns^2$ electronic 
configuration ($n = 2, 3, 4, 5$, and $6$) is close to the excited state 
with the $p$ electrons, 
the ground state of $ns^2$ electronic configuration repels with 
the excited state with $np$ electrons 
very strongly in the perturbation theory through 
the mixing between the $ns^2$ and charge-excited configuration 
in covalent and polarizable materials, 
then the charge excitation dramatically stabilizes the electron-rich 
configuration in $ns^2$ electronic configuration than the $ns^1$ 
electronic configuration.  
Therefore, we can observe universally in the experiment that 
the $ns^1$ electronic configuration always shows 
CEI-NEU and the missing oxidation state in the thermal equilibrium 
(naturally it does not exist in the nature).  

Based on CEI-NEU system, 
we can design an itinerant system with NEU 
$s^1$ electronic configuration 
in alkali metal [for example 
Li$^0$($2s^1$), 
Na$^0$($3s^1$), 
K$^0$($4s^1$), 
Rb$^0$($5s^1$), and 
Cs$^0$($6s^1$)], 
or + charged ($s^1$) state 
in alkaline-earth metal [for example 
Be$^+$($2s^1$), 
Mg$^+$($3s^1$), 
Ca$^+$($4s^1$), 
Sr$^+$($5s^1$), 
Zn$^+$($4s^1$), 
Cd$^+$($5s^1$), 
Ba$^+$($6s^1$), and 
Hg$^+$($6s^1$)] 
by applying the ultra-high pressures in order to increase 
a charge density at the atomic site with reducing the volume about 
50\%$\sim$70\%.  
Typical examples are Li$^0$($2s^1$) and Ca$^+$($4s^1$) 
which show the superconductivity at 20 K for 
Li and 25 K for Ca under the ultra-high pressures~\cite{12, 13}.
Other candidates for the itinerant 
and CEI-NEU system are 
summarized in Table I. 

It is easy to fabricate transparent conducting oxides (TCO) 
in $n$-type doped ZnO or In$_2$O$_3$ upon the donor doping such as 
oxygen-vacancy (double donors), or applied the ultra-high pressures, 
therefore, these system has high potentiality to realize 
the transparent high-$T_c$ superconductors upon the $n$-type doping or 
under the conditions of the ultra-high pressures. 

Here, we propose a $T_c$-enhancement in NEU layers 
for high-$T_c$ superconductivity, 
which are constructed by (i) the $T_c$-enhancing layers 
increasing the pairing interaction by 
CEI-NEU system (or atoms, molecules, or clusters) 
and (ii) the covalent-superconducting layers with a strong covalent-bonding 
with resisting the formation of CDW, or SDW caused 
by the strong NEU system.

When we look at the electronic configurations of ground states 
in $3d$ transition atoms, 
we see the missing state in the $3d^4$ electronic configuration 
in Cr$^0$ atom, where the $3d^54s^1$ electronic configuration 
becomes more stable than the $3d^44s^2$ electronic configuration 
due to the exchange-correlation energy gain. 
When a free atom is placed into the polarizable host materials such as 
oxides or semiconductors, its Coulomb and exchange-correlation interactions 
respond in fundamentally different ways to screening.  
The Coulomb interaction, responding to long-wave length (mono-pole) screening, 
is reduced far more than the exchange-correlation interaction 
(multi-pole screening)~\cite{14}.  
This has been demonstrated theoretically and experimentally~\cite{15}, 
as is called Haldane and Anderson mechanism~\cite{16}.  
Therefore, for $3d$ transition atoms in the condensed matter, 
we can expect the ECI-NEU 
in $d^4$ (or $d^6$) electronic configuration, 
where the large exchange-correlation energy-gain stabilizes 
the $d^5$ electronic configurations than the $d^4$ (or $d^6$) 
electronic configuration, resulting in the following 
charge disproportionation or dynamical charge fluctuation 
by a missing oxidation states in $d^4$ (or $d^6$) electronic configurations 
through the reaction of the $2d^4 \rightarrow d^3 + d^5$ ($U<0$), 
[or $2d^6 \rightarrow d^5 + d^7$  ($U<0$)].  
The candidates of ECI-NEU system 
are Cr$^{2+}$($3d^4$), Mn$^{3+}$($3d^4$), Fe$^{4+}$($3d^4$),
Co$^{5+}$($3d^4$), Ni$^{6+}$($3d^4$) for $d^4$, and 
Cr$^0$($3d^6$), Mn$^+$($3d^6$), Fe$^{2+}$($3d^6$), 
Co$^{3+}$($3d^6$), and Ni$^{4+}$($3d^6$) for $d^6$ 
configurations (see Table II.)

In the ground state configuration, we can also see the missing oxidation state 
in the $3d^9$ electronic configuration in Cu$^0$ atom, 
where the $3d^{10}4s^1$ electronic configuration becomes 
more stable than the $3d^94s^2$ electronic configuration 
due to the charge-excitation-induced energy gain through 
the charge excitation from $3d^{10}4s^1$. 
Due to the same reason as was discussed in ECI-NEU, 
its long-range Coulomb and short-range Coulomb interactions 
respond in fundamentally different ways to screening, 
where a long-range Coulomb interaction responding to screening 
is reduced far more than the short-range $d$-$s$ charge-excitation interaction. 
Thus $d^9$ electronic configuration such as 
Cu$^{2+}$($3d^9$) and Ag$^{2+}$($4d^9$) is 
CEI-NEU system through the stabilization energy of 
electron-rich $3d^{10}$ (closed-shell stabilization) 
electronic configuration. 
Therefore we can expect the superconductivity 
caused by the pairing through 
CEI-NEU with dynamical charge disproportionation 
(or dynamical charge fluctuation) 
if we can avoid anti-ferromagnetic SDW upon $p$- or $n$-type doping 
with the itinerant system.  
For cuprate high-$T_c$ superconductors, 
such as La$_{2(1-x)}$Sr$_{2x}$CuO$_4$ or YBa$_2$Cu$_3$O$_7$, 
the ground state of undoped system is anti-ferromagnetic and 
charge-transfer insulator caused by the super-exchange interactions 
in the CuO$_2$ layer. 
It may be possible that the superconductivity is caused (or enhanced) 
by CEI-NEU in the Cu$^{2+}$ 
($3d^9$) in the itinerant system by destabilizing 
the anti-ferromagnetic SDW ordering upon $p$-type doping 
such as La$_{2(1-x)}$Sr$_{2x}$CuO$_4$, 
where we may see the crossover of the anti-ferromagnetic SDW 
by super-exchange interaction and the superconductivity caused by 
CEI-NEU in Cu$^{2+}$
($3d^9$) system.  
The possibility of NEU-induced superconductivity 
caused by charge-fluctuation-induced NEU is proposed and discussed 
by Varma, combined with the missing oxidation states 
in BaPb$_{1-x}$Bi$_x$O$_3$ and BK$_{1-x}$BiO$_3$~\cite{17}. 

For the cuprate high-$T_c$ superconductors such as 
La$_{2(1-x)}$Sr$_{2x}$CuO$_4$, YBa$_2$Cu$_3$O$_7$, 
Bi$_2$Sr$_2$Ca$_2$Cu$_3$O$_{10}$, (Hg,Re)Ba$_2$CaCu$_2$O$_y$, 
HgBa$_2$Ca$_2$Cu$_3$O$_8$, Tl$_2$Ba$_2$CaCu$_2$O$_8$, 
we may have an additional-pairing mechanism in the $T_c$-enhancing layers 
which is caused by the CEI-NEU 
in $s^1$ electronic configuration in the electron-doped 
BaO, SrO, BiO, TlO, Y$_2$O$_3$, PbO or HgO layers upon 
the oxygen-vacancy doping.   
For new-type superconductors such as MgB$_2$~\cite{29}, CaSi$_2$~\cite{30}, 
it is also possible to expect an additional pairing mechanism for 
$T_c$-enhancement caused by the CEI-NEU 
in Mg$^+$($3s^1$) or Ca$^+$($4s^1$) as a result of 
electron condensation at the atomic site originated from $\pi$-orbital 
perpendicular to the B or Si layers.  
Even, it is also possible to enhance the pairing mechanism 
by CEI-NEU in the $s^1$ 
electronic configurations of K$^0$($4s^1$), Rb$^0$($5s^1$), 
Cs$^0$($6s^1$) or Be$^+$($2s^1$), Mg$^+$($3s^1$), Ca$^+$($4s^1$), 
Sr$^+$($5s^1$), Zn$^+$($4s^1$), Cd$^+$($5s^1$), Ba$^+$($6s^1$), 
and Hg$^+$($6s^1$), with partial ionization caused by the $\pi$-orbital 
hybridization perpendicular to the sphere or cluster surfaces 
in the conducting-covalent large molecules or clusters, 
such as C$_{60}$, Zeolite, or, 
[Ca$_{24}$Al$_{28}$O$_{64}$]$^{4+}$($4e^-$). 

Recently, it was discovered a new-type high-Tc superconductors 
in the Ni- and Fe-based oxypnictides, such as 
La(O$_{1-x}$F$_x$)FeP($T_c = 5$ K)~\cite{21}, 
LaONiP ($T_c = 3$ K)~\cite{21}, 
La(O$_{1-x}$F$_x$)FeAs ($T_c = 26$ K)~\cite{22}, 
GdO$_{1-x}$FeAs ($T_c = 53.5$ K)~\cite{23}, 
NdO$_{0.6}$FeAs ($T_c = 55$ K)~\cite{24}, 
Sm(O$_{1-x}$F$_x$)FeAs ($T_c = 43$ K)~\cite{25}, 
La$_{0.85}$Sr$_{0.13}$FeAs ($T_c = 25$ K)~\cite{26}, 
LaO$_{1-x}$F$_x$FeAs ($T_c = 46$ K at $P=4$ GPa)~\cite{27}, 
and (Ba$_{1-x}$K$_x$)Fe$_2$As$_2$ ($T_c = 38$ K)~\cite{28}.  
These new-type superconductors also contain 
CEI-NEU ions, such as 
La$^{2+}$($6s^1$), Nd$^{2+}$($6s^1$), Gd$^{2+}$($6s^1$), 
Sm$^{2+}$($6s^1$), with the electron-doped LaO, GdO, NdO, 
and SmO-layers by F-donor doping or oxygen-vacancy-donor doping. 
Based upon our materials design as discussed above, 
we have a possibility to enhance the $T_c$ by using another layers 
of NEU ions as 1+ ions in $n$-type doped 
ZnO, CaO, MgO, BeO, BaO, CdO, SrO, HgO, In$_2$O$_3$, Al$_2$O$_3$, 
B$_2$O$_3$, Y$_2$O$_3$, Ga$_2$O$_3$, Tl$_2$O$_3$, La$_2$O$_3$, 
CeO$_2$, Pr$_2$O$_3$, Eu$_2$O$_3$, etc., or 
$p$-type doped Ce$_2$O$_3$, etc., or 3+ ions in $n$-type doped 
CO$_2$, SiO$_2$, TiO$_2$, ZrO$_2$, GeO$_2$, SnO$_2$, PbO$_2$, 
etc., or in $p$-type doped GeO, SnO, PbO, etc. or 
4+ ions in $n$-type doped VO$_2$, NbO$_2$, As$_4$O$_{10}$, 
NaAsO$_3$, Ab$_4$O$_{10}$, Bi$_2$O$_5$, and NaBiO$_3$, etc., 
or $p$-type doped VO$_2$, NbO$_2$, LiNbO$_2$, As$_2$O$_3$, Sb$_4$O$_6$, 
and Bi$_2$O$_3$, etc. 
These are all promising candidates for 
$T_c$-enhancing layers by CEI-NEU. 

The oxypnictides contain Fe$^{2+}$($3d^6$) which is 
ECI-NEU system as was discussed above. 
We should also try the candidates of ECI-NEU 
system such as Cr$^{2+}$($3d^4$), Mn$^{3+}$($3d^4$), 
Fe$^{4+}$($3d^4$), Co$^{5+}$($3d^4$), Ni$^{6+}$($3d^4$) for 
$d^4$, and Cr$^0$($3d^6$), Mn$^+$($3d^6$), Co$^{3+}$($3d^6$), 
Ni$^{4+}$($3d^6$) for $d^6$ configurations (see Table II.)

We design a new-class of NEU system for 
the realization of new-class of high-$T_c$ superconducting materials 
by controlling (i) CEI-NEU 
in $s^1$ (or $d^9$) electronic configurations, 
and (ii) ECI-NEU in $d^4$ 
(or $d^6$) electronic configurations.  
Here, we have proposed the general rules and chemical trends of 
NEU system for the realization of high-$T_c$ 
superconductivity in real materials.  
We also proposed the materials design for the realization of 
high-$T_c$ superconductivity based on the general rules and 
chemical trends comparing with the available experimental data.

\acknowledgement
This work was partially supported 
by Grant-in-Aid for Scientific Research 
from the Japan Society for the Promotion of Science 
and the Ministry of Education, Culture, Sports, Science and 
Technology (MEXT), 
Global Center of Excellence Program by MEXT, 
New Energy and Industrial Technology Development Organization Program and 
Japan Science and Technology Agency Program.

\begin{figure}[htbp]
\begin{center}
\includegraphics[width=0.8\hsize]{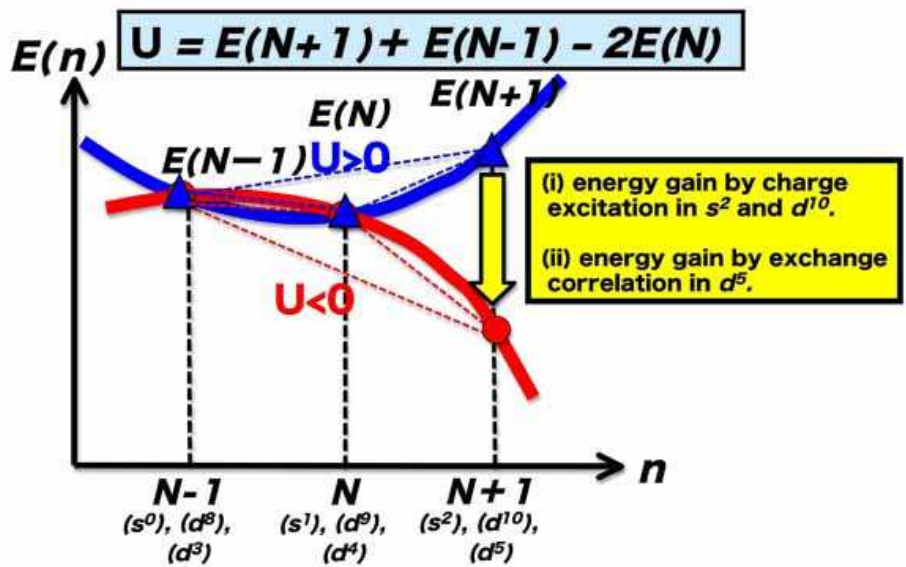}
\caption{
Definition of effective correlation energy $U$ 
($U= E(N+1) + E(N-1) - 2E(N)$ ) of $N$ electron system, 
where $E(N)$ is the total energy of $N$ electron system.  
Normally, positive effective $U$ ($U>0$) is realized due to 
the repulsive Coulomb interaction between two electrons 
[indicating the convexity in $E(N)$ as a function of $N$].  
Negative effective $U$ ($U<0$ ) occurs, 
if $N+1$, or, $N-1$ electron system becomes more stable than 
the $N$ electron system [indicating the concavity in $E(N)$], 
which is caused by (i) the charge-excitation energy gain or 
(ii) exchange-correlation energy gain.}
\label{fig1}
\end{center}
\end{figure}

\begin{table}[htbp]
\caption{
The candidates for the $T_c$-enhancing layers in the superconductivity, 
where $s^1$ electronic configurations indicate negative effective $U$. 
}
\label{t2}
\begin{tabular}{cll}
\hline
Charge & Negative effective $U$ center & Materials \\
\hline
+1 &  Zn$^+$($4s^1$), Ca$^+$($4s^1$), Mg$^+$($2s^1$), Be$^+$($2s^1$), &
$n$-type doped ZnO, CaO, MgO, \\
& Cd$^+$($5s^1$), Ba$^+$($6s^1$), Sr$^+$($5s^1$), Hg$^+$($6s^1$) &
BeO, CdO, BaO, SrO, HgO \\
\hline
+2 & In$^{2+}$($5s^1$), Al$^{2+}$($3s^1$), B$^{2+}$($2s^1$), Y$^{2+}$($5s^1$), &
$n$-type doped In$_2$O$_3$, Al$_2$O$_3$, B$_2$O$_3$, \\
& Ga$^{2+}$($4s^1$), Tl$^{2+}$($6s^1$), La$^{2+}$($6s^1$), Ce$^{2+}$($5s^1$), &
Y$_2$O$_3$, Ga$_2$O$_3$, Tl$_2$O$_3$, La$_2$O$_3$, \\
& Pr$^{2+}$($5s^1$), Nd$^{2+}$($6s^1$), Sm$^{2+}$($6s^1$), &
CeO$_2$, Pr$_2$O$_3$, Nd$_2$O$_3$, Sm$_2$O$_3$, \\
& Eu$^{2+}$($6s^1$), Gd$^{2+}$($6s^1$)  & 
Eu$_2$O$_3$, Gd$_2$O$_3$, \\
&& $p$-type doped Ce$_2$O$_3$ \\
\hline
+3 & C$^{3+}$($2s^1$), Si$^{3+}$($3s^1$), Ti$^{3+}$($4s^1$), 
Zr$^{3+}$($5s^1$), & 
$n$-type doped CO$_2$, SiO$_2$, \\
&&TiO$_2$, ZrO$_2$, \\
& Ge$^{3+}$($4s^1$), Sn$^{3+}$($5s^1$), Pb$^{3+}$($6s^1$) &
GeO$_2$, SnO$_2$, PbO$_2$, \\
&&$p$-type doped GeO, SnO, PbO \\
\hline
+4 & N$^{4+}$($2s^1$), P$^{4+}$($3s^1$), V$^{4+}$($4s^1$), Nb$^{4+}$($5s^1$), &
$n$-type doped VO$_2$, NbO$_2$, As$_4$O$_{10}$, \\
&As$^{4+}$($4s^1$), Sb$^{4+}$($5s^1$) Bi$^{4+}$($6s^1$)&
NaAsO$_3$, Sb$_4$O$_{10}$, Bi$_2$O$_5$, NaBiO$_3$, \\
&&$p$-type doped VO$_2$, NbO$_2$, LiNbO$_2$, \\
&&As$_2$O$_3$, Sb$_4$O$_6$, Bi$_2$O$_3$\\
\hline
+5 & O$^{5+}$($2s^1$), S$^{5+}$($3s^1$), 
Cr$^{5+}$($4s^1$), Mo$^{5+}$($5s^1$), &
$n$-type doped MoO$_3$, SeO$_2$, TeO$_3$, \\
& Se$^{5+}$($4s^1$), Te$^{5+}$($5s^1$), Po$^{5+}$($6s^1$)&
$p$-type doped MoO$_2$, H$_2$SeO$_4$, TeO$_2$ \\
\hline
+6 &
F$^{6+}$($2s^1$), Cl$^{6+}$($3s^1$), 
Mn$^{6+}$($4s^1$), Tc$^{6+}$($5s^1$), & 
$n$-type doped H$_5$IO$_6$, HIO$_4$, \\
&& KBrO$_4$, MnO$_7$, HClO$_4$, \\
& Br$^{6+}$($4s^1$), I$^{6+}$($5s^1$), At$^{6+}$($6s^1$) &
$p$-type doped I$_2$O$_3$, HIO$_3$, \\
&&NBrO$_3$, HCl$_2$O$_3$\\
\hline
\end{tabular}
\end{table}

\begin{table}[htbp]
\caption{
(a) Exchange-correlation-induced negative effective $U$ ions 
for $d^4$ and $d^6$ electronic configuration, 
and (b) charge-excitation-induced negative effective $U$ ions 
for $d^9$ electronic configuration.
}
\label{t3}
\ \\
(a) 
\begin{tabular}{|c|c|c|c|c|}
\hline 
2+ & 3+ & 4+ & 5+ & 6+ \\ \hline
Cr$^{2+}$ ($3d^4$) &
Mn$^{3+}$ ($3d^4$) &
Fe$^{4+}$ ($3d^4$) &
Co$^{5+}$ ($3d^4$) &
Ni$^{6+}$ ($3d^4$) \\ \hline
0 & + & 2+ & 3+ & 4+ \\ \hline
Cr$^{0}$ ($3d^6$) &
Mn$^{+}$ ($3d^6$) &
Fe$^{2+}$ ($3d^6$) &
Co$^{3+}$ ($3d^6$) &
Ni$^{4+}$ ($3d^6$) \\ \hline
\end{tabular}
\ \\
(b) 
\begin{tabular}{|c|}
\hline
2+ \\ \hline
Cu$^{2+}$ ($3d^9$) \\ \hline
Ag$^{2+}$ ($3d^9$) \\ \hline
\end{tabular}
\end{table}





\end{document}